# A novel computational technique using coefficient diagram method for load frequency control in an interconnected power system

*Jalal Heidary [a*], Hassan Rastegar[a]*

[a]*Department of Electrical Engineering, Amirkabir University of Technology, Tehran, Iran*

**Abstract**

This paper proposes a novel load frequency control (LFC) approach formulated on an optimal structure of the coefficient diagram method (CDM) in a two-area thermal power system. As part of a realistic analysis, nonlinearities related to governor dead band (GDB) and generation rate constraint (GRC) have been considered. In this article, a hybrid CDM method is combined with the optimization of its mathematical equations to achieve an innovative controller. Furthermore, a new metaheuristic optimization technique called the water cycle algorithm (WCA) is used to determine the optimal coefficients of the CDM controller. For the purpose of demonstrating the validity of the proposed scheme, a wide range of uncertainties in the dynamic parameters of a nonlinear power system were tested. In addition, a comparative study is presented between the results obtained from a classical integral, CDM alone, optimized Fuzzy, optimized PID, and the suggested controller. In this new approach to improved control, algebraic support provides a robust and responsive controller that can provide fast and stable dynamic responses and effectively overcome the detrimental effects of nonlinearities and uncertainties in the parameters of the power system.

# 1. Introduction

Within the power system network, frequency control is one of the challenges that must be addressed properly. It is becoming more important because of uncertainties and the complexity of the electrical network [1]. The early aim of LFC is to maintain the scheduled system frequency and inter-area tie-line power within predetermined limits to cope with the load variations and system disturbances [2,3]. Control strategies that are robust, easy to implement, and take into account the growing complexity and variation in the power system are required to achieve desirable performance.

Over the last decades, control system designers have applied several research and design strategies to the LFC problem to find the best solutions. In LFC problems, Proportional Integral Derivative (PID) is still a suitable controller. Conventional techniques for determining PID parameters need a linearized mathematical model of the system under study. Recently, heuristic and metaheuristic optimization techniques like Genetic [4], Cuckoo search [5], Bacterial foraging [6], Imperialist competitive [7], Whale optimization algorithm [8], and chaotic optimization [9] have been discussed to overcome these issues. In Ref [10], load frequency control for the power system with uncertainties such as GDB and GRC has been investigated. Control systems that employ fixed parameters, such as PID controllers, are incapable of handling uncertainties and load perturbations due to their design at specific nominal operating points. Therefore, robust and adaptive control strategies in the papers address this problem since these traditional control strategies may no longer perform well in all operating conditions.

Fuzzy logic is one of these control approaches to solve the LFC problem. Unlike conventional methods that rely primarily on a mathematical model for analysis, Fuzzy logic control is based on experience and knowledge of experts about a system. [11]. The success of fuzzy logic controllers is also presented in Ref. [12-15]. In these papers, several optimizations techniques have been carried out to find the gains of the fuzzy controllers. Ref. [12] uses a sine cosine algorithm, and Ref. [13] presents a Fuzzy tuning PI controller using a Self-Modified Bat Algorithm (SAMBA) for tuning the parameters. Also, Type-2 fuzzy controller [14] and Fractional-order Fuzzy PID [15-17] have been introduced to enhance the performance of the Fuzzy approach in the LFC.

Other techniques such as neural network in Ref. [18-19], H∞ controller, sliding mode controller, and Predictive Control in Ref. [20-24] have been applied for the LFC problem. By implementing these strategies, the performance of dynamic controllers is improved when dealing with nonlinearities and uncertainties compared to conventional methods. However, complex calculations to obtain parameters and difficulties in physical implementation continue to obstruct their use in power system grids. Considering these concerns, and with the power system structure becoming more complex day by day, new intelligence techniques are essential to provide desirable performance for power systems.

Recently, a new method of load control has been introduced based on coefficient diagrams. The method uses an algebraic approach to indirectly determine the poles and settle time by using a polynomial overall closed loop [25]. While CDM has a relatively recent application in LFC, its primary principle has been investigated in other fields. Servo motors, steel mill drives, gas turbines, and spacecraft attitude control are examples of these applications. CDM is a polynomial technique that has been developed to guarantee the robustness of the solution. From the shape of the diagram, the designer can see the response, stability, and robustness. Therefore, the design procedure is easier to understand and less complicated [25].

In Ref. [26], a LFC method using CDM is proposed. In this paper, comparing this CDM controller with integrators and model predictive controllers, its superiority has been demonstrated. However, the coefficient and parameters to tune the controller have not been optimized. In Ref. [27], state feedback gains and Kalman filters were used to improve the CDM controller performance. Despite the fact that this article presents a superior control technique, state feedback-based controllers are not practical because they are too complex and are not available in real power systems. In Ref. [28], rather than merely using the classical technique of decentralizing algebraic CDM, the suggested control scheme in this research work takes into account an optimization technique to illustrate the best performance of this controller. Despite the fact that this paper presents a new approach to finding the best coefficients for CDM controllers, it has a lot of tunable parameters, making it complicated, while the CDM has the potential to demonstrate proper performance with much fewer variables.

Along with all the many benefits of the CDM controller, which have already been presented in the mentioned research works, this paper introduces a new design approach to find the best performance of this controller with respect to its algebraic equations. Moreover, to make the controller practical for industrial applications, the proposed CDM controller's tuning parameters have been decreased compared to previous papers. The key parameters of the proposed controller have been determined using an optimization algorithm called the water cycle algorithm (WCA). The WCA is a new optimization algorithm that has been proposed by the observation of the water cycle process and the movements of rivers and streams toward the sea [29]. The performance of WCA in tuning CDM parameters compared to GA and PSO has also been investigated in the present work. A variety of cases, including large steps and uncertainty in dynamic parameters, were used to illustrate the superiority of the proposed method. In this study, comparing the results of using a classical integral, a CDM alone, a fuzzy strategy, an optimized PID, and the proposed controller, the superiority of the suggested controller has been confirmed. The objectives of this work are:

(a) Optimize the parameters of CDM with respect to its algebraic equations.
(b) Using an optimization technique called WCA with better performance to find the coefficients of CDM controller.

(c) Making the CDM Controller more practical by minimizing the number of tunable parameters.

(d) Comparing the proposed control strategy with existing ones to demonstrate its superiority.

(e) Show the robustness of the suggested controller against uncertainties and large disturbances.

This paper is organized as follows: "Dynamic response of power system" provides an overview of the power system structure and its essential control loops. In the "coefficient diagram method" section, a general explanation of this method is described. The "Water Cycle Algorithm" section discussed a new optimization technique used in this paper. In "proposed control strategy", the CDM-OPT method is introduced. Several simulations for studying the performance of the control strategy and the results are in section "Results and simulations." The paper ends with the "conclusion" section, followed by the references.

## 2. Dynamic response of power system

An extensive interconnected power system with multiple generation types includes a number of different areas [30]. High-voltage transmission lines interconnect these areas. Fig. 1 shows a controlled multi-area power system with primary droop control and secondary supplementary feedback loops [1]. Primary control is not fast enough to bring the local frequency back to its steady-state point. Therefore, the secondary control loop delivers reserve power to decrease the frequency deviations [31]. A load variation in each area affects the system's response in all areas; therefore, a fast and robust controller is necessary to improve frequency fluctuation in the power system.

The total power variation in tie-line between area-i and the other areas is as follows:

$$\Delta Ptie, i = \frac{2\pi}{s}\left(\sum_{\substack{j=1\\j\neq i}}^{N} T_{ij}\Delta f_i - \sum_{\substack{j=1\\j\neq i}}^{N} T_{ij}\Delta f_j\right) \quad (1)$$

The deviation in frequency represented as:

$$\Delta \dot{f}_i = \left(\frac{1}{2H_i}\right).\Delta Pm_i - \left(\frac{1}{2H_i}\right).\Delta PL_i - \left(\frac{D_i}{2H_i}\right).\Delta f_i - \left(\frac{1}{2H_i}\right).\Delta Ptie, i \quad (2)$$

In addition, the dynamics of the turbine calculated as:

$$\Delta \dot{P}_{mi} = \left(\frac{1}{T_{ti}}\right).\Delta Pg_i - \left(\frac{1}{T_{ti}}\right).\Delta P_{mi} \quad (3)$$

Moreover, the dynamic of the governor represented by:

$$\Delta \dot{P}_{gi} = \left(\frac{1}{T_{gi}}\right).\Delta Pc_i - \left(\frac{1}{R_i T_{gi}}\right).\Delta f_i - \left(\frac{1}{T_{gi}}\right).\Delta Pg_i \quad (4)$$

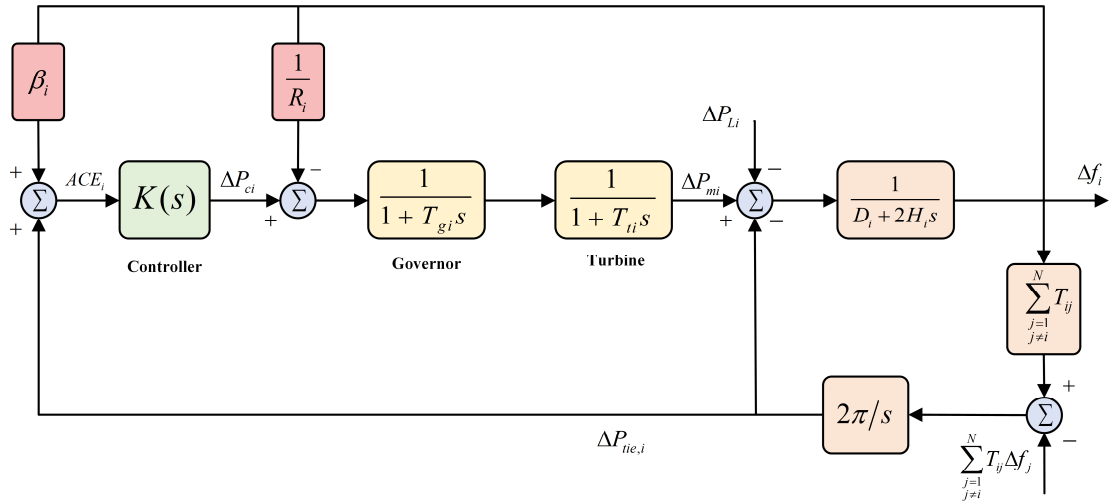

**Fig. 1.** Block diagram schematic of i<sup>th</sup> area power system used in the study [26].

## 3. Coefficient diagram method

The CDM is a polynomial design method, where instead of a Bode diagram, the coefficient diagram is used, and the sufficient condition for stability by Lipatov forms its theoretical basis [32]. CDM provides information on the stability, time response, and robustness characteristic of the system under the study in a single diagram [26].

As shown in Fig. 2, a two-parameter configuration is used to implement an overall transfer function in CDM design. $N(s)=B_p(s)$ is numerator polynomial, $D(s)=A_p(s)$ is dominator polynomial of the system. In addition, $F(s)$ is the reference numerator while $A_c(s)$ and $B_c(s)$ are forward denominator and feedback numerator polynomials respectively [33].

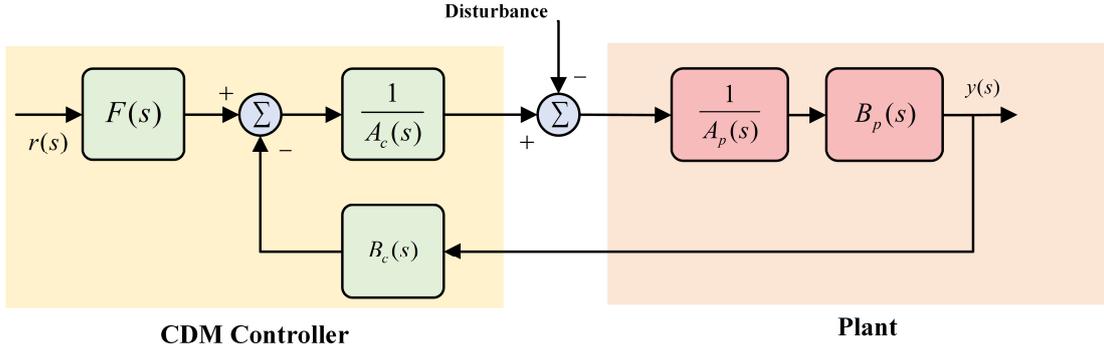

**Fig. 2.** Schematic diagram of a CDM controller.

Note that subscript c and p stand for controller and plant, respectively.

In this approach, $r(s)$ is considered as the reference input to the system, $d$ as the external disturbance and $u$ as the control signal. The controller's output, $y$, can be determined by:

$$y = \frac{N(s)F(s)}{P(s)}r(s) + \frac{N(s)A_c(s)}{P(s)}d \tag{5}$$

where $P(s)$ is the characteristic polynomial of the closed-loop system, and it is defined by:

$$P(s) = A_c(s)A_p(s) + B_c(s)B_p(s) = a_n s^n + \cdots + a_1 s + a_0 = \sum_{i=0}^{n} a_i s^i \tag{6}$$

The design parameters of CDM controllers are the equivalent time constant ($\tau$), the stability indices ($\gamma_i$), and the stability limits ($\gamma^*$). The relations between these parameters and the coefficients of the characteristic polynomial ($a_i$) can be described as follows:

$$\gamma_i = \frac{a_i^2}{a_{i+1} \cdot a_{i-1}} \quad , i = 1,2,\ldots,(n-1), \gamma_0 = \gamma_i = \infty \tag{7}$$

$$\tau = \frac{a_1}{a_0} \tag{8}$$

$$\gamma_i^* = \frac{1}{\gamma_{i-1}} + \frac{1}{\gamma_{i+1}} \quad , i = 1,2,\ldots,(n-1), \gamma_0 = \gamma_i = \infty \tag{9}$$

The designer, as per the controller's requirement, can change the above $\gamma_i$ values. However, proper tuning of these parameters requires experience or trial and error.

Also, define the pseudo-break points

$$\omega_i = \frac{a_i}{a_{i-1}} \quad , i = 1,2,\ldots,(n-1) \tag{10}$$

Then we have

$$\gamma_i = \frac{\omega_i}{\omega_{i-1}} \quad , \quad i = 1,2,\ldots,(n-1) \tag{11}$$

$$a_i = \frac{a_0 \tau^i}{\gamma_{i-1} \gamma_{i-2}^2 \cdots \gamma_2^{i-2} \gamma_1^{i-1}} \tag{12}$$

Considering the parameters ($\tau$) and ($\gamma_i$) the objective characteristic polynomial, $P_{target}(s)$ can be expressed as follows:

$$P_{target}(s) = a_0 \times \left[\left\{\sum_{i=2}^{n}\left(\prod_{j=1}^{i-1}\frac{1}{\gamma_{i-j}^j}\right)(\tau s)^i\right\} + (\tau s) + 1\right] \tag{13}$$

Moreover, the reference numerator polynomials $F(s)$ can be determined from:

$$F(s) = \left(\frac{P(s)}{N(s)}\right)\bigg|_{s=0} \tag{14}$$

## 4. Water cycle algorithm

Inspired by the water cycle process in nature and how rivers and streams flow towards the sea, a population-based metaheuristic algorithm, the Water Cycle Algorithm (WCA), has recently been introduced [34]. Similar to the other metaheuristic approach, the suggested algorithm begins with an initial population of design variables called streams that are randomly generated from the raining in nature [35]. In an $N_{Var}$ dimensional optimization problem, a stream is an array of $1 \times N_{Var}$. Consider this array as:

$$A\ candidate\ stream = [x_1, x_2, x_3, \ldots, x_N] \tag{15}$$

To start the optimization, a candidate demonstrating a matrix of streams of size $N_{pop} \times N_{Var}$ that is generated randomly, defined as follows:

$$Total\ population = \begin{bmatrix} sea \\ River_1 \\ River_2 \\ \vdots \\ Stream_{N_{sr}+1} \\ Stream_{N_{sr}+2} \\ \vdots \\ Stream_{N_{pop}} \end{bmatrix} = \begin{bmatrix} x_1^1 & x_2^1 & \cdots & x_N^1 \\ x_1^2 & x_2^2 & \cdots & x_N^2 \\ \vdots & \vdots & \vdots & \vdots \\ x_1^{N_{pop}} & x_2^{N_{pop}} & \cdots & x_N^{N_{pop}} \end{bmatrix} \tag{16}$$

The intensity of flow for each stream is obtained by the evaluation of cost function (C) given as:

$$C_i = Cost_i = f(x_1^i, x_2^i, \ldots, x_{N_{var}}^i) \quad i = 1,2,3, \ldots, N_{pop} \tag{17}$$

where $N_{pop}$ and $N_{Var}$ are the numbers of streams (initial population) and the number of tunable variables, respectively. For the first step, $N_{pop}$ streams are generated. A number of $N_{sr}$ from the best individuals are chosen as a sea and rivers. The stream that has the minimum value among others is considered as a sea. In fact, $N_{sr}$ is the summation of the number of rivers and a single sea as given in Eq. (18). The rest of the population (streams that flow to the rivers or may alternatively directly flow to the sea) is computed using the following equation:

$$N_{sr} = Number\ of\ Rivers + 1 \tag{18}$$

$$N_{Raindrops} = N_{pop} - N_{sr} \tag{19}$$

In order to designate streams to the rivers and sea, the following equation is given:

$$NS_n = round\left\{\left|\frac{cost_n}{\sum_{i=1}^{N_{sr}} cost_i}\right| \times N_{Raindrops}\right\}, \quad n = 1,2,\ldots,N_{sr} \tag{20}$$

where $NS_n$ is the number of streams, which flow to the particular rivers or sea. According to the following formula, the stream flows to the river along the line connecting them:

$$X \in (0, C \times d), \quad C > 1 \tag{21}$$

where $1 < C < 2$ and the best value for C may be selected as 2. The current distance between stream and river is represented as $d$ [34]. The exploitation phase in WCA, the new position for streams and rivers can be given as:

$$X_{stream}^{i+1} = X_{stream}^i + rand \times C \times (X_{River}^i - X_{Stream}^i) \tag{22}$$

$$X_{stream}^{i+1} = X_{stream}^i + rand \times C \times (X_{sea}^i - X_{stream}^i) \tag{23}$$

$$X_{River}^{i+1} = X_{River}^i + rand \times C \times (X_{sea}^i - X_{River}^i) \tag{24}$$

where rand is a random number between 0 and 1. If the solution obtained by a stream is better than its connecting river, the positions of the river and stream are switched (i.e., stream becomes river and river becomes stream). A similar interchange can be happened for a river and the sea and for the streams and the sea as well.

In addition, the evaporation and raining process are suggested in WCA to enhance its performance. To be more precise, evaporation can prevent the algorithm from rapid convergence and is responsible for the exploration phase in the WCA.

$$if \ |X_{Sea}^i - X_{River}^i| < d_{max} \quad i = 1,2,3,\ldots,N_{sr} - 1 \tag{25}$$

*Evaporation and raining process end*

where $d_{max}$ is a small number (close to zero). It can control the search intensity near the sea (the optimum solution). The value of $d_{max}$ decreases as follows [34]:

$$d_{max}^{i+1} = d_{max}^{i} - \frac{d_{max}^{i}}{\max iteration} \tag{26}$$

For determining the positions of the newly created streams, the following equation is applied:

$$X_{Strean}^{new} = LB + rand \times (UB - LB) \tag{27}$$

where *LB* and *UB* are lower and upper limits defined by the problem, respectively. The flowchart for WCA is outlined in Sect. A.4 of Appendix Section.

## 5. Proposed control strategy

Although the standard CDM is simple and proven to be robust, there may be some challenges involved in implementing it as a controller for the power system, particularly for LFC. A major challenge in power system frequency control is the fact that the complexity of power system structure and the frequent changes within a power system make it difficult to tune CDM parameters. [25]. Obtaining the key parameters ($\gamma_i, \tau$) for the CDM controller is the vital step in the process of designing the controller. Previous papers chose these variables without utilizing any optimization method despite having a set of mathematical equations. Moreover, previous papers used different stability indices and time constants for each area, resulting in increasing the number of tunable parameters. However, this paper attempts to resolve these problems by applying the optimal CDM controller with a limited number of parameters to the load frequency control problem. This innovative controller uses a CDM technique that is well integrated with algebraic optimization throughout its equations. The stability indices ($\gamma_i$), time constant ($\tau$) and ($K_{B0}$) are considered as the tuning parameters of CDM controller in this study. It should be noted these parameters were determined in Ref [26] without any optimization.

$$CDM\ tune\ parametrs:\quad \gamma_i:[\gamma_0,\gamma_1,\gamma_2,\ldots]\,,\tau\,,K_{B0} \tag{28}$$

In order to make a more robust controller with fewer variables, only $K_{B0}$ is different in each area, while the same stability indices, and time constant are considered for both areas of the power system. After calculating *N(s), D(s)* of the plant transfer function and generation new populations of CDM parameters by an optimization algorithm, *A<sub>c</sub>(s)* and *B<sub>c</sub>(s)* can be calculated by equations (13) and (16).

In order to find the best coefficient for *A<sub>c</sub>(s)* and *B<sub>c</sub>(s)* an objective function should be defined. In this study, integral absolute error (IAE) of the frequency deviation of all areas is chosen as the objective function. The objective function J is set to be:

$$J = IAE = \int_0^\infty (|\Delta f_i|)dt \tag{29}$$

Note that subscript i stands for CDM controller in area i. In order to optimize the gains of the CDM controller; a 1% step increment in loads of all areas with a 50% change in governor and turbine parameters is applied. Fig. 3 shows the flow chart of the proposed method in this study.

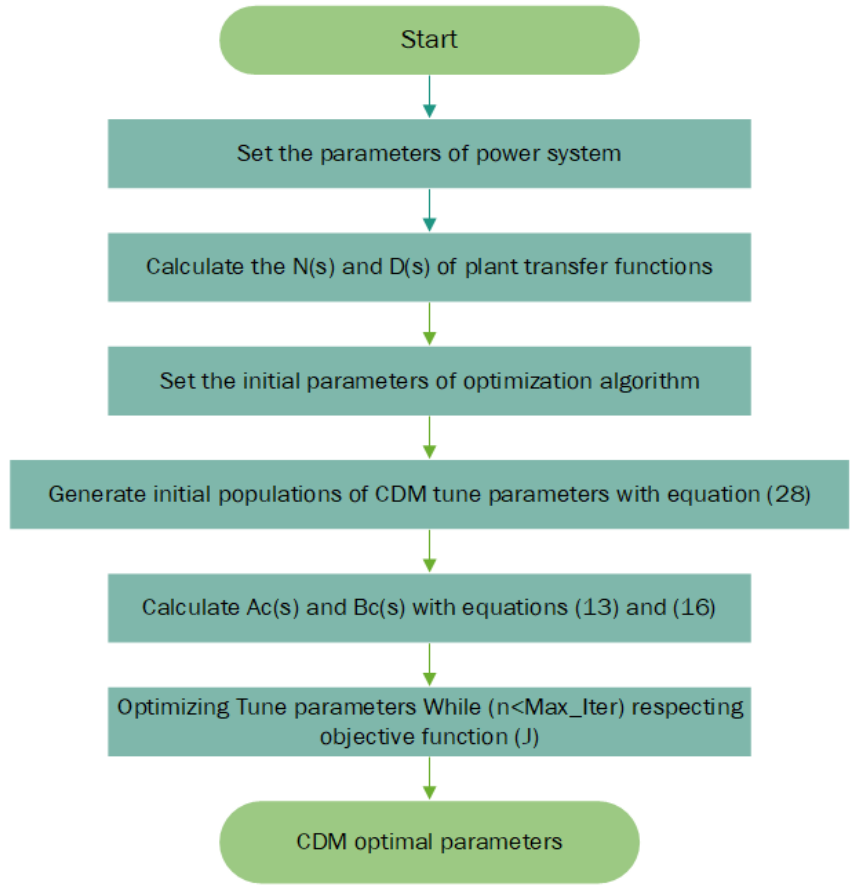

**Fig. 3.** Flow chart of procedure to tune CDM controller.

As a test system, a two-area nonlinear power system has been chosen in order to compare different control strategies applied to load frequency control [26]. The parameters of the power systems are shown in Table1. The major observations of the proposed method are presented below. For each of this power system's areas, the GRC has been set at 10% per minute and GDB as 0.05 pu.

**Table 1.**
Two-area power system network [26].

| Area | D (pu/Hz) | 2H (pu/sec) | R(Hz/pu) | $T_g$ (sec) | $T_t$(sec) | $T_{12}$(pu/Hz) |
|---|---|---|---|---|---|---|
| 1 | 0.015 | 0.1667 | 3 | 0.08 | 0.4 | 0.2 |
| 2 | 0.016 | 0.2017 | 2.73 | 0.06 | 0.44 | |

Related algebraic calculations in order to find these CDM optimal parameters are given. Table 2 summarizes these equations and compares CDM-OPT characteristics with classical CDM.

*Area 1*

$N_1(s) = 0.3483S+1.256$

$D_1(s) = 0.005334 S^4+0.805 S^3+0.1739 S^2+0.348 S$

$P_{target,1} = 0.000000732 S^6+0.001 S^5+0.054 S^4+0.276 S^3+0.0793 S^2+2.276 S+2.57$

$\gamma_i = [25.33, 0.01, 17.62, 9.88, 29.98],\ \tau = 0.8832\ \ K_{B0,1} = 20.5126$

$A_1(s) = 2.0318S+0.0014S^2$

$B_1(s) = 20.5126+12.4314S+6.9261S^2$

*Area 2*

$N_2(s) = 0.3827S+1.256$

$D_2(s) = 0.00532 S^4+0.10127 S^3+0.2097 S^2+0.382 S$

$P_{target,2} = 0.000001789 S^6+0.0026 S^5+0.133 S^4+0.673 S^3+0.193 S^2+5.540 S+6.27$

$\gamma_i = [25.33, 0.01, 17.62, 9.88, 29.98],\ \tau = 0.8832\ \ K_{B0,2} = 39.9347$

$A_2(s) = 3.9521S+0.0027S^2$

$B_2(s) = 39.9347+23.1225S+11.917S^2$

**Table 2.**
CDM controller with and without WCA optimum values for two-area power system.

| Method | area | τ | $K_{B0}$ | Stability indices | A(s) | B(s) |
|---|---|---|---|---|---|---|
| CDM-OPT | area 1 | 0.8832 | 20.5126 | [25.33, 0.01 17.62, 9.88, 29.98] | $2.0318S+0.0014S^2$ | $20.5126+12.4314S+6.9261S^2$ |
|  | area 2 |  | 39.9347 |  | $3.9521S+0.0027S^2$ | $39.9347+23.1225S+11.917S^2$ |
| CDM [26] | area 1 | 2 | 40 | [1, 6.5, 1.5, 2.7, 72] | $150S+2S^2$ | $40+69S+100S^2$ |
|  | area 2 |  | 32 | [1, 6.4, 2.3, 1.53, 3.5] | $60S+3S^2$ | $32+54S+100S^2$ |

# 6. Results and simulations

In this section, various comparative cases are examined to illustrate the effectiveness and robustness of the proposed strategy. Simulation of a two-area power system, considering nonlinearities such as GRC and GDB based on the proposed strategy, has been done using MATLAB/SIMULINK toolbox.

### 6.1 Case1: Convergence of optimization techniques.

According to the discussion in section 5, the parameters of the CDM controller can be tuned by different optimization algorithms. In this study, the WCA is chosen to make the CDM controller show its best performance. In order to investigate the effectiveness of the WCA, a comparison has been carried out with the GA and PSO. The conditions for the simulations are:

1- A 1% step increase in load demand of both areas at t=1sec and t=30sec.

2- The best performance of each algorithm is chosen from ten runs.

3- According to strategy control in section 5, the objective function J is set to be:

$$J = IAE = \int_0^t |\Delta f_1|dt + \int_0^t |\Delta f_2|dt \qquad (30)$$

The other three performance indices to investigate the responses are:

$$ISE = \int_0^t \Delta f_1^2 dt + \int_0^t \Delta f_2^2 dt \qquad (31)$$

$$ITSE = \int_0^t (\Delta f_1^2 + \int_0^t \Delta f_2^2) t dt \qquad (32)$$

$$ITAE = \int_0^t (|\Delta f_1| + \int_0^t |\Delta f_2|) t dt \qquad (33)$$

Assuming the same number of initial population (chosen as 50) and the max iteration (chosen as 50), the comparative convergence of profile of objective function, as shaped by different algorithms, has been carried out. The characteristics of these algorithms are presented in the Sect. A.4 of Appendix. The performance of each algorithm independently is shown in Fig. 4(a), Fig. 4(b), and Fig. 4(c), respectively. Fig. 4(d) compares the best performance of each algorithm together. It can be observed that WCA exhibits a faster convergence profile. Table

3. shows the max, min, average, and standard deviation of the final values of the objective functions gathered in these 30 simulations (10 simulations for each optimization). It can be seen that WCA offers a smaller final of objective function. Another advantage of WCA is the simulation time that is decreased by 20 percent. Thus, WCA may be apprehended as an effective optimization tool for tuning the parameters of CDM controllers in LFC applications. So, only WCA has been considered for the rest of the work.

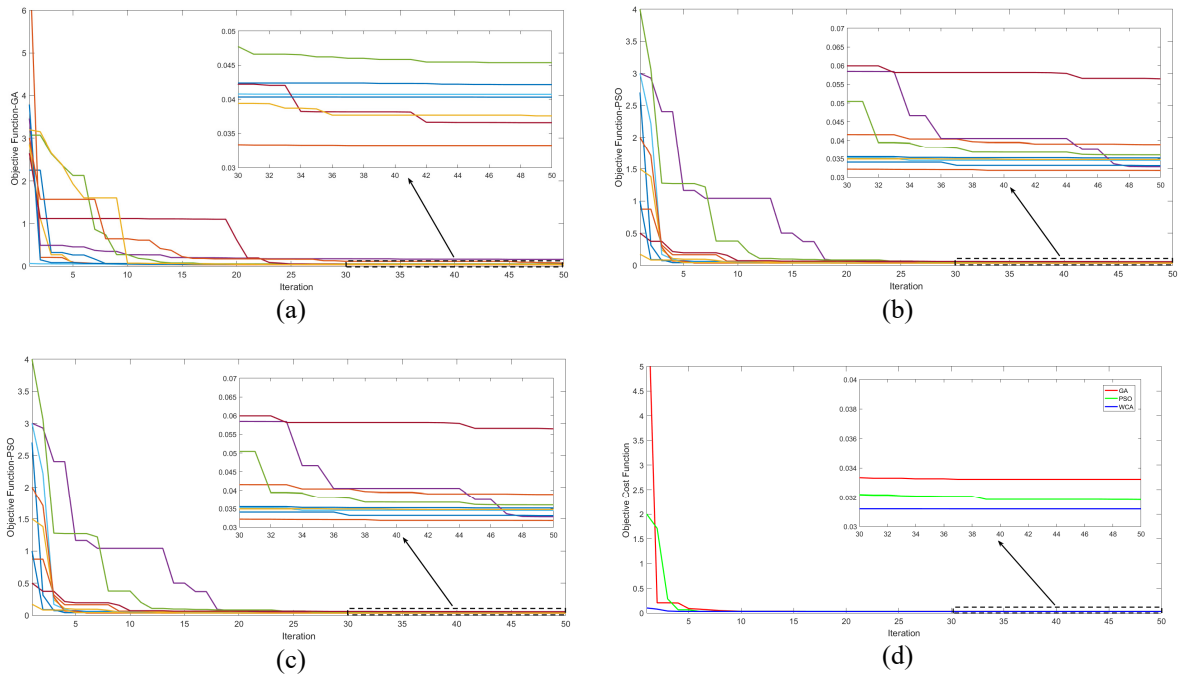

**Fig. 4.** Case 1 convergence profiles: (a) GA. (b) PSO. (c) WCA. (d) Best minimum objective function.

**Table 3.**
Statistical analysis of objective functions.

| Algorithm | Min | Max | Average | Standard deviation |
|:---:|:---:|:---:|:---:|:---:|
| GA | 0.03323 | 0.15970 | 0.0562 | 0.0381 |
| PSO | 0.03184 | 0.05647 | 0.0368 | 0.0072 |
| WCA | 0.03120 | 0.03635 | 0.0336 | 0.0018 |

*6.2 Case 2: Step load demand change in area 1*

Five control strategies have been considered to compare the performance of the controllers. Parameters of these controllers are presented in Sect. A.2 and A.3 of Appendix. Following is a list of these five control methods:

1- Integral controller [1].

2- Classical CDM controller [26].

3- Optimized PID controller with WCA [35].

4- Optimized Fuzzy controller with WCA[17].

5- Optimized CDM ( Proposed method)

As the second case, a 1% step increase in load demand of the area-1 is applied. Fig. 5(a) Shows frequency change in area 1, Fig. 5(b) shows frequency change in area 2, and Fig. 5(c) shows tie-line power variations between the areas. The various performance measures of system responses have been tabulated in Table 4 for each controller. In addition, due to the small OS, US, and decreasing the level of oscillations compared to other strategies, the proposed method leads to better performance.

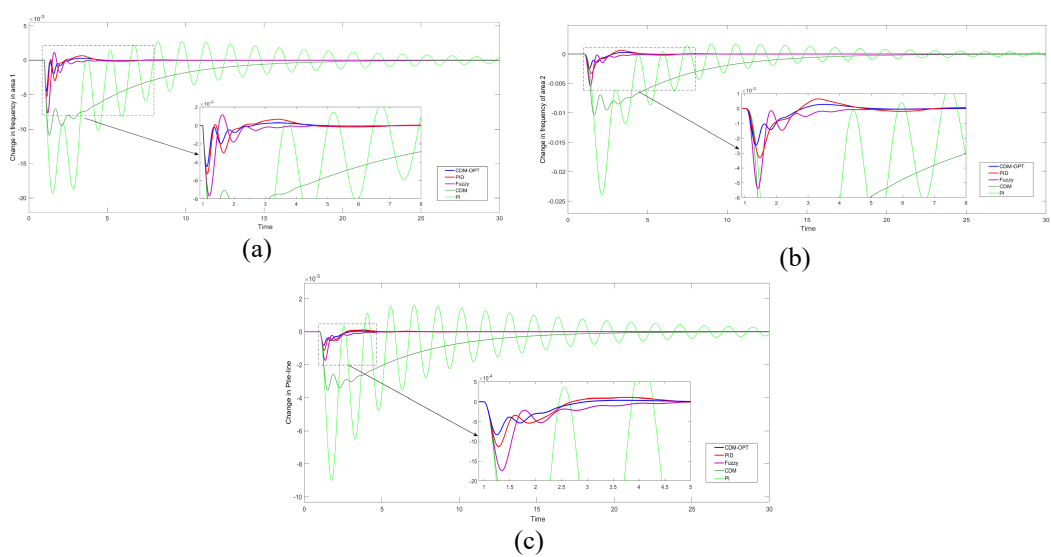

**Fig. 5.** Case 2 signals: (a) Frequency in area-1. (b) Frequency in area-2. (c) Tie-line between areas.

**Table 4.**
System response for step change in load, pertaining to case 2.

| Method | $\Delta f_1$ | | | | $\Delta f_2$ | | | | $\Delta P_{tie}$ | | | |
|---|---|---|---|---|---|---|---|---|---|---|---|---|
| | $t_s(t)$ | US | OS | IAE(×10⁻³) | $t_s(t)$ | US | OS | IAE(×10⁻³) | $t_s(t)$ | US | OS | IAE(×10⁻³) |
| *CDM-OPT* | 6.71 | 2.907e-04 | -4.508e-03 | 2.179 | 6.55 | 2.711e-04 | -2.480e-03 | 2.083 | 3.87 | 0.3647e-04 | -0.8345e-03 | 0.677 |
| *PID* | 8.41 | 6.549e-04 | -5.287e-04 | 3.278 | 9.7 | 6.394-04 | -3.313e-03 | 3.161 | 6.42 | 1.032e-04 | -1.131e-03 | 0.960 |
| *Fuzzy* | 6.5 | 11.50e-04 | 7.701e-04 | 3.631 | 6.33 | N.O | -5.381e-03 | 3.246 | 3.87 | N.O | -1.741e-03 | 0.687 |
| *CDM* | 30> | N.O | -10.89e-03 | 55.08 | 30> | N.O | -10.42e-03 | 55.07 | 30> | N.O | -3.544e-03 | 20.12 |
| *PI* | 30> | 27.3e-04 | -19.33e-03 | 73.25 | 30> | 17.25e-04 | -23.97e-03 | 63.47 | 30> | 16.25e-04 | -9.005e-03 | 31.39 |

*6.3 Case 3: Sinusoidal load change*

In this case, a sinusoidal load disturbance is applied to area-1. Fig. 6(a) shows the load variations in area-1. Fig. 6(b) shows frequency change in area 1, Fig. 6(c) shows frequency change in area 2, and Fig. 6(d) shows tie-line power variations between the areas. It is evident that by providing fast response, the optimized CDM controller dampens oscillations to zero faster than other methods.

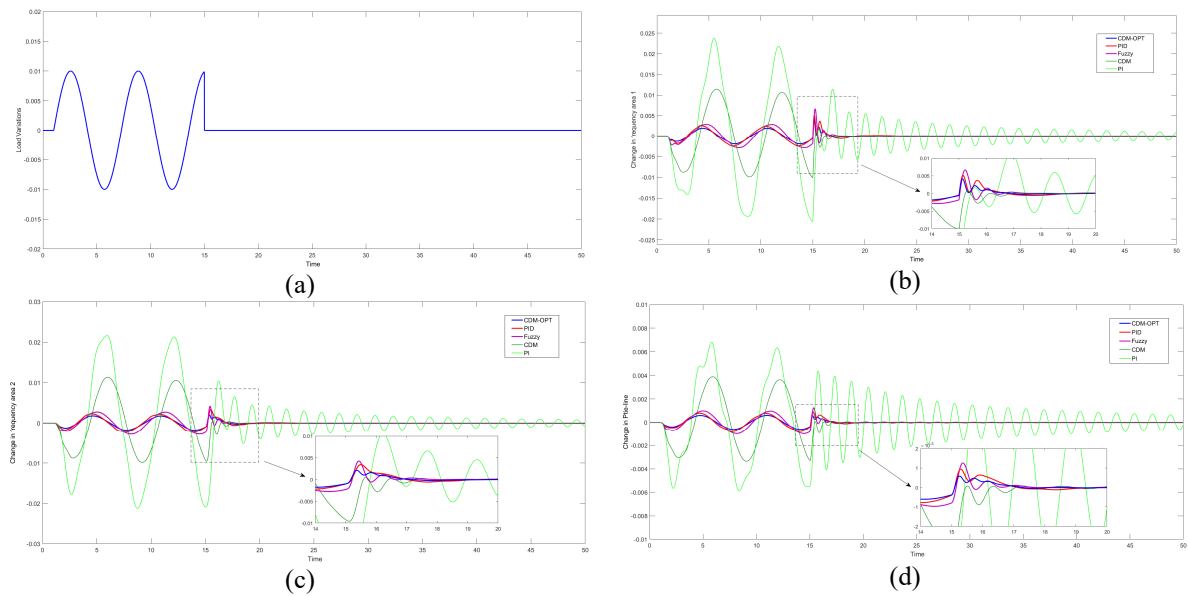

**Fig. 6.** Case 3 Signals: (a) Sinusoidal load variation. (b) Frequency in area-1. (c) Frequency in area-2. (d) Tie-line between areas

*6.4 Case 4: Uniformly distributed random load*

In this case, a random load disturbance is applied to both area-1 and area-2. Fig. 7(a) and Fig. 7(b) show the load variations in these areas. Frequency deviations of area-1, area-2, and tie-line power are plotted in Fig. 7(c), Fig. 7(d), and Fig. 7(e), respectively. It is evident that the designed CDM-OPT controller shows a better performance compared to other controllers.

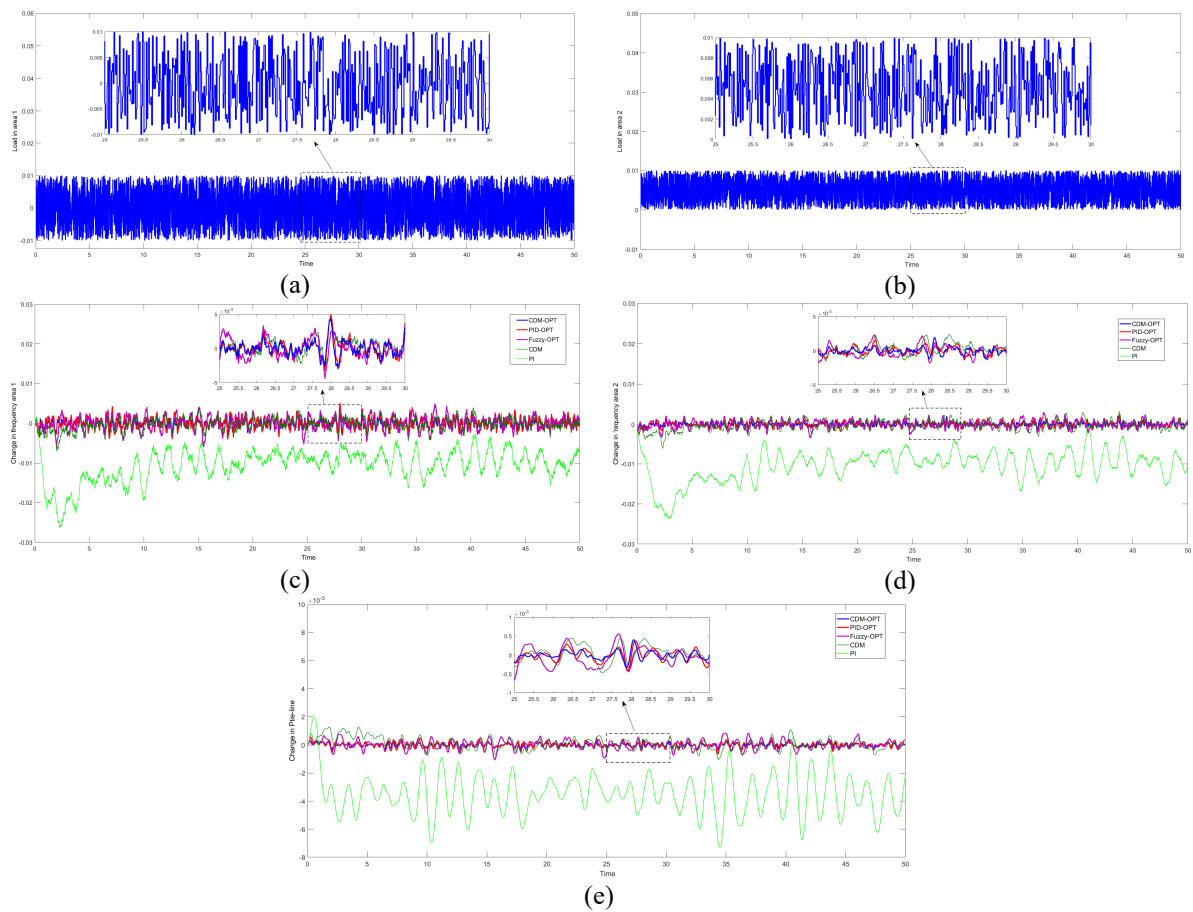

**Fig. 7.** Case 4 Signals: (a) Random load in area-1. (b) Random load in area-2. (c) Frequency in area-1. (d) Frequency in area-2. (e) Tie-line between areas.

*6.5 Case 5: Load variations in all areas with changes in system parameters*

In order to examine the effectiveness of the proposed method, the system is tested against a wide range of uncertainties and random load demands in both areas altogether. The governor time constants increased to $T_{g1}$=0.105 (31% change), $T_{g2}$=0.105 (66% change) and turbine time constants increased to $T_{t1}$=0.785 (95% change), $T_{t2}$=0.6 (38% change) respectively. Fig. 8a shows the load variations, and Fig. 8b, Fig. 8c and Fig. 8d show other related responses of the controllers. Table 5 shows a comparison analysis between cases 2-5. It has been indicated that the designed controller can provide sufficient response under different operating conditions.

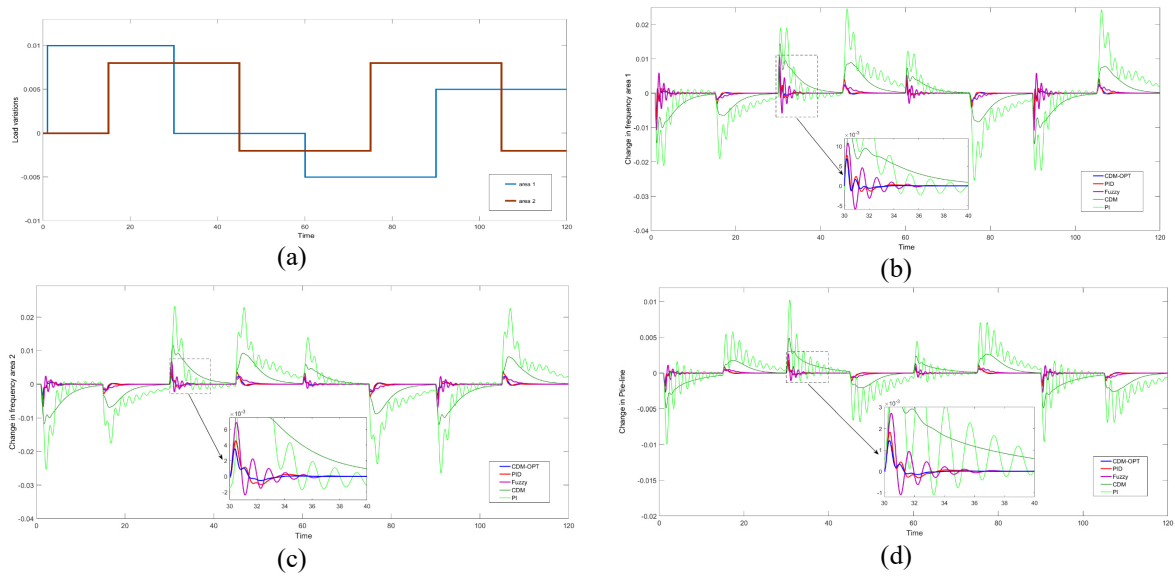

**Fig. 8.** Case 5 Signals: (a) load variations. (b) Frequency in area-1. (c) Frequency in area-2. (d) Tie-line between areas.

**Table 5.**
Responses analysis in cases 2-5.

| Case | Indices | Controller | | | | |
|---|---|---|---|---|---|---|
| | | CDM-OPT | PID | Fuzzy | CDM | PI |
| Case 2 | ISE | $6.422 \times 10^{-6}$ | $11.52 \times 10^{-6}$ | $23.37 \times 10^{-6}$ | $611.7 \times 10^{-6}$ | $1248 \times 10^{-6}$ |
| | ITSE | $9.192 \times 10^{-6}$ | $18.11 \times 10^{-6}$ | $32.06 \times 10^{-6}$ | $2395 \times 10^{-6}$ | $4149 \times 10^{-6}$ |
| | ITAE | $9.113 \times 10^{-3}$ | $16.62 \times 10^{-3}$ | $12.97 \times 10^{-3}$ | $661.3 \times 10^{-3}$ | $1110 \times 10^{-3}$ |
| | IAE (J) | $4.262 \times 10^{-3}$ | $6.439 \times 10^{-3}$ | $6.887 \times 10^{-3}$ | $110.3 \times 10^{-3}$ | $141.3 \times 10^{-3}$ |
| Case 3 | ISE | $4.994 \times 10^{-5}$ | $8.506 \times 10^{-5}$ | $12.05 \times 10^{-5}$ | $147.5 \times 10^{-5}$ | $609.5 \times 10^{-5}$ |
| | ITSE | $4.758 \times 10^{-4}$ | $8.272 \times 10^{-4}$ | $11.53 \times 10^{-4}$ | $124.5 \times 10^{-4}$ | $587.6 \times 10^{-4}$ |
| | ITAE | $3.514 \times 10^{-1}$ | $4.777 \times 10^{-1}$ | $5.097 \times 10^{-1}$ | $16.56 \times 10^{-1}$ | $62.06 \times 10^{-1}$ |
| | IAE (J) | $3.624 \times 10^{-2}$ | $4.803 \times 10^{-2}$ | $5.435 \times 10^{-2}$ | $18.89 \times 10^{-2}$ | $4745 \times 10^{-2}$ |
| Case 4 | ISE | $5.763 \times 10^{-5}$ | $9.967 \times 10^{-5}$ | $18.52 \times 10^{-5}$ | $19.26 \times 10^{-5}$ | $1260 \times 10^{-5}$ |
| | ITSE | $1.391 \times 10^{-3}$ | $2.372 \times 10^{-3}$ | $4.249 \times 10^{-3}$ | $3.527 \times 10^{-3}$ | $249.4 \times 10^{-3}$ |
| | ITAE | 1.403 | 1.878 | 2.498 | 2.339 | 23.89 |
| | IAE (J) | $5.718 \times 10^{-2}$ | $7.654 \times 10^{-2}$ | $10.45 \times 10^{-2}$ | $10.69 \times 10^{-2}$ | $105.2 \times 10^{-2}$ |
| Case 5 | ISE | $8.438 \times 10^{-5}$ | $13.98 \times 10^{-5}$ | $30.33 \times 10^{-5}$ | $418.4 \times 10^{-5}$ | $1352 \times 10^{-5}$ |
| | ITSE | $4.336 \times 10^{-3}$ | $7.28 \times 10^{-3}$ | $14.18 \times 10^{-3}$ | $226.3 \times 10^{-3}$ | $819.1 \times 10^{-3}$ |
| | ITAE | 2.371 | 3.031 | 5.587 | 39.72 | 74.04 |
| | IAE (J) | $4.409 \times 10^{-2}$ | $5.786 \times 10^{-2}$ | $10.65 \times 10^{-2}$ | $70.7 \times 10^{-2}$ | $121.0 \times 10^{-2}$ |

*6.6 Case 6: Sensitivity analysis*

An analysis of sensitivity is performed to determine how robust the proposed optimized CDM controller is to a wide range of system dynamic parameters. A 1% step increase in load demand of the area-1 is applied. Fig. 9 shows the objective function value with $\pm 25\%$, $\pm 50\%$ variations in governor and turbine parameters at the same time. Table 6 shows the sensitivity analysis for governor and turbine time constant in each area separately. Fig. 10(a) and Fig. 10(b) demonstrate the robustness of the controller against other power system uncertainties. Analyzing the sensitivity of the results suggested the control scheme provided complete robustness, stability, and fast response in all simulated scenarios, regardless of the dynamics.

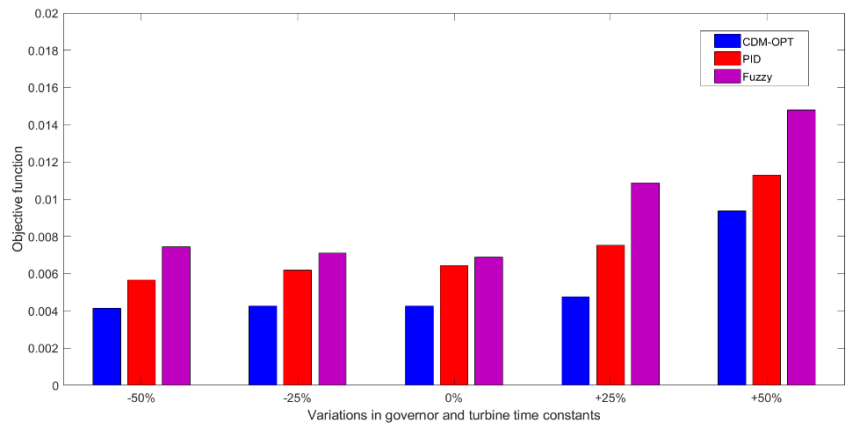

**Fig. 9.** Objective function value with $\pm 25\%$, $\pm 50\%$ variations in governor and turbine parameters.

**Table 6.**
Sensitivity analysis for turbine and governor time constant (ISE, ITSE (×10⁻⁶), IAE, ITAE (×10⁻³)).

| Time constant | Variation %Age | Value | CDM-OPT ISE | ITSE | ITAE | IAE (J) | PID ISE | ITSE | ITAE | IAE(J) | Fuzzy ISE | ITSE | ITAE | IAE(J) |
|---|---|---|---|---|---|---|---|---|---|---|---|---|---|---|
| *Nominal* | - | - | 6.422 | 9.192 | 9.113 | 4.262 | 11.52 | 18.11 | 16.62 | 6.439 | 23.37 | 32.06 | 12.97 | 6.877 |
| $T_{g1}$ | +25% | 0.1 | 7.311 | 10.26 | 8.856 | 4.236 | 13.18 | 20.25 | 16.48 | 6.565 | 27.75 | 38.84 | 14.36 | 7.544 |
|  | -25% | 0.06 | 5.845 | 8.646 | 9.669 | 4.380 | 10.46 | 16.88 | 16.85 | 6.485 | 20.42 | 28.17 | 13.51 | 6.839 |
|  | +50% | 0.12 | 8.576 | 12.02 | 8.999 | 4.465 | 15.97 | 24.71 | 17.05 | 7.061 | 33.94 | 55.34 | 12.23 | 12.78 |
|  | -50% | 0.04 | 5.472 | 8.327 | 9.892 | 4.429 | 9.863 | 16.38 | 17.11 | 6.552 | 18.67 | 26.48 | 14.18 | 7.113 |
| $T_{g2}$ | +25% | 0.075 | 6.510 | 9.312 | 8.999 | 4.240 | 11.65 | 18.25 | 16.50 | 6.395 | 23.56 | 32.32 | 13.02 | 6.875 |

| | -25% | 0.045 | 6.350 | 9.110 | 9.294 | 4.295 | 11.38 | 17.95 | 16.75 | 6.487 | 23.26 | 31.95 | 12.93 | 6.889 |
| --- | --- | --- | --- | --- | --- | --- | --- | --- | --- | --- | --- | --- | --- | --- |
| | +50% | 0.09 | 6.605 | 9.456 | 8.987 | 4.233 | 11.86 | 18.58 | 16.53 | 6.406 | 23.94 | 32.96 | 13.38 | 6.996 |
| | -50% | 0.03 | 6.316 | 9.109 | 9.506 | 4.350 | 11.25 | 17.80 | 16.86 | 6.526 | 23.21 | 31.91 | 12.89 | 6.906 |
| $T_{t1}$ | +25% | 0.5 | 7.961 | 11.11 | 9.568 | 4.507 | 14.62 | 22.66 | 17.81 | 6.990 | 30.23 | 41.61 | 13.35 | 7.459 |
| | -25% | 0.3 | 5.334 | 7.988 | 9.321 | 4.194 | 9.536 | 15.61 | 16.27 | 6.281 | 17.82 | 25.11 | 13.92 | 6.778 |
| | +50% | 0.6 | 9.742 | 13.45 | 10.29 | 4.820 | 18.53 | 28.86 | 19.76 | 7.854 | 38.07 | 53.28 | 17.98 | 8.802 |
| | -50% | 0.2 | 4.935 | 7.722 | 9.582 | 4.225 | 8.284 | 13.92 | 15.45 | 6.014 | 13.86 | 20.98 | 15.28 | 7.020 |
| $T_{t2}$ | +25% | 0.55 | 6.535 | 9.386 | 9.311 | 4.302 | 11.73 | 18.22 | 16.22 | 6.339 | 23.93 | 32.95 | 12.15 | 6.727 |
| | -25% | 0.33 | 6.279 | 8.962 | 9.130 | 4.244 | 11.21 | 17.63 | 16.59 | 6.435 | 23.04 | 31.59 | 13.23 | 6.978 |
| | +50% | 0.66 | 6.633 | 9.523 | 9.273 | 4.307 | 12.32 | 19.46 | 16.92 | 6.610 | 24.83 | 34.69 | 12.78 | 6.923 |
| | -50% | 0.22 | 6.176 | 8.877 | 9.457 | 4.316 | 10.78 | 16.85 | 16.26 | 6.348 | 22.95 | 31.59 | 13.74 | 7.148 |

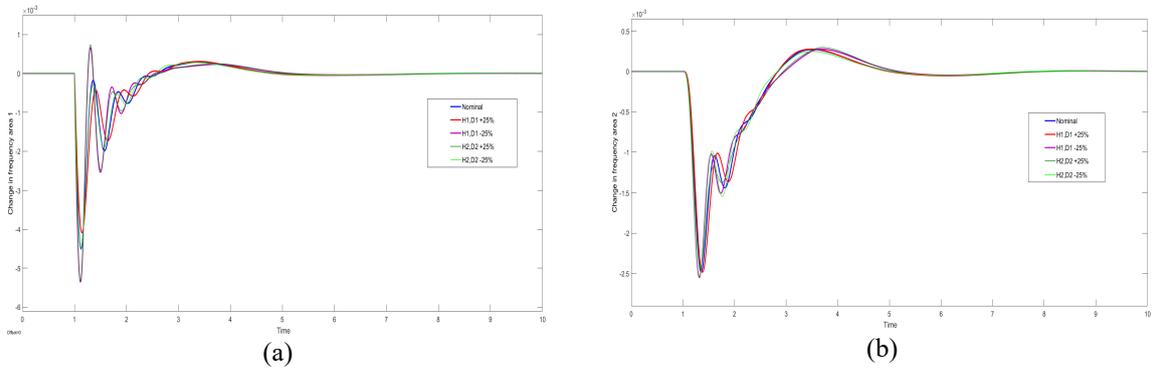

**Fig. 10.** (a) Frequency deviations of area-1. (b) Frequency deviations of area-2.

## 7. Conclusion

For frequency control of a thermal two-area power system, in which nonlinearities, including GRC and GDB, are taken into consideration, this paper developed a mathematically comprehensible technique for optimizing variables of the coefficient diagram method controller. Innovative aspects of the suggested controller include the combination of the CDM technique and optimization within its mathematical equations. In addition, WCA is suggested as the optimization technique to tune the parameters of the CDM controller, and its performance regarding minimum fitness value and convergence rate has been investigated compared to GA and PSO. Furthermore, the suggested strategy combines the tunable parameters of the CDM controller for each area, decreasing the number of variables by using the same stability indices for both areas to avoid further complexity, making it much easier to implement. A step load disturbance, sinusoidal load perturbations, random loads, as well as scenarios dealing with uncertainty have all been used to verify the effectiveness of the proposed scheme. A comparison was made between classical integral, CDM alone, fuzzy optimized, PID optimized results, and suggested optimized CDM controller. In all cases, system frequency deviation is properly controlled, and the CDM-OPT controller provides better response respecting to undershoot, overshoot, settling time, and several performance indices, which confirmed the superiority of the suggested control over other methods.

# Appendix

*(A.1) Fuzzy controller* [17].

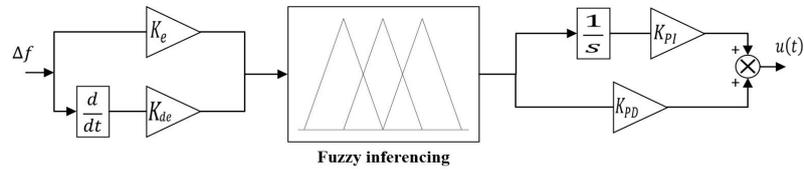

**Fig .A.1** Schematic of the Fuzzy controller.

TableA.1
Rule base for Fuzzy controller.

| e $\frac{d}{dt}$ | NL | NM | NS | ZR | PS | PM | PL |
|---|---|---|---|---|---|---|---|
| PL | ZR | PS | PM | PL | PL | PL | PL |
| PM | NS | ZR | PS | PM | PL | PL | PL |
| PS | NM | NS | ZR | PS | PM | PL | PL |
| ZR | NL | NM | NS | ZR | PS | PM | PL |
| NS | NL | NL | NM | NS | ZR | PS | PM |
| NM | NL | NL | NL | NM | NS | ZR | PS |
| NL | NL | NL | NL | NL | NM | NS | ZR |

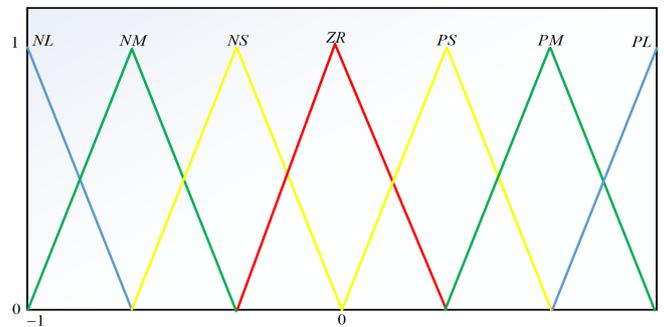

**Fig .A.2** Membership functions for inputs and outputs of the Fuzzy controller.

**TableA.2**
WCA optimal values of Fuzzy controller.

| Parameter | two area power system | |
|---|---|---|
| | area 1 | area 2 |
| $K_e$ | 0.7514 | 0.4311 |
| $K_{de}$ | 0.3033 | 0.3074 |
| $K_{PI}$ | 3.3655 | 6.0457 |
| $K_{PD}$ | 2.1726 | 7.9261 |

*(A.2) PID controller*[34].

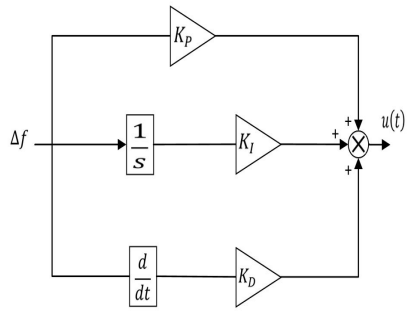

**Fig .B.1** Schematic of the PID controller.

**TableB.1**
WCA optimal value of PID controller and integral controller value.

| Parameter | two area power system | |
|---|---|---|
| | area 1 | area 2 |
| $K_P$ | 3.8830 | 4.4420 |
| $K_I$ | 8.9908 | 8.1478 |
| $K_D$ | 2.9089 | 1.0651 |
| Integral | 0.3 | 0.2 |

*(A.3) Optimization algorithms:*

Tune parameters:

**WCA**: $N_{pop}$=50, Max-it=50, $N_{sr}$=4, dmax=$10^{-16}$ (Default values for WCA [34])

**PSO**: $N_{pop}$=50, Max-it=50, $C_1$= $C_2$=1.49 (Default values, MATLAB function= *particleswarm*)

**GA**= $N_{pop}$=50, Max-it=50, Crossover=0.8 (Default values, MATLAB function= *GA*)

*Note: None of the tunable parameters of algorithms were modified to illustrate a fair comparison between default values.*

Flowchart of WCA [34].

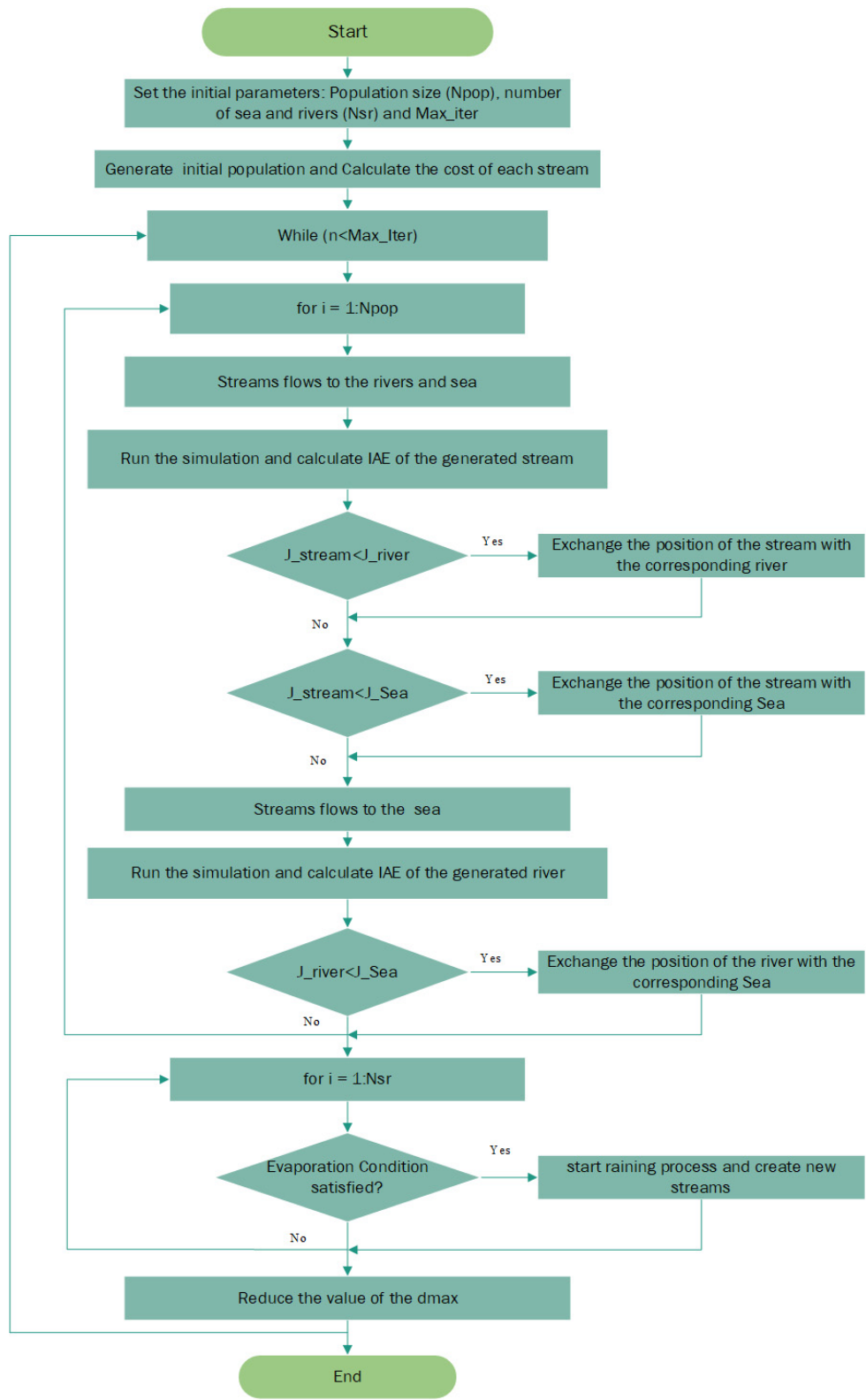

# List of Abbreviations

| | |
|---|---|
| LFC | Load frequency control |
| GDB | Governor dead band |
| GRC | Generation rate constraint |
| CDM | Coefficient diagram method |
| WCA | Water cycle algorithm |
| GA | Genetic algorithm |
| PSO | Particle swarm optimization |
| IAE | Integral of squared error |
| ISE | Integral of time multiplied absolute error |
| ITAE | Integral of time multiplied squared error |
| ITSE | Integral of time multiplied squared error |
| $ACE$ | Area control error |
| $\Delta P_{tie,i}$ | Total tie-line power variation between area i and the other areas |
| $T_{ij}$ | Tie-line synchronizing coefficient between area i and j |
| $\Delta f_i$ | Frequency deviation of area i |
| $\Delta P_g$ | Governor output |
| $\Delta P_m$ | Mechanical power change |
| $\Delta P_c$ | Control power |
| $\Delta P_L$ | Load change |
| $T_{gi}$ | Governor time constant |
| $T_t$ | Turbine time constant |
| $H$ | Equivalent inertia constant |
| $D$ | Damping constant |
| $R$ | Droop characteristics |
| $B$ | Frequency bias coefficient |
| $B_p(s)$ | Numerator of polynomial of the plant transfer function |
| $A_p(s)$ | Denominator of polynomial of the plant transfer function |
| $A_c(s)$ | Forward denominator polynomial of the plant transfer function |
| $B_c(s)$ | Feedback numerator polynomial |
| $r(s)$ | Reference input |
| $F(s)$ | Reference numerator polynomial |
| $P(s)$ | Characteristic polynomial of the closed-loop system |
| $\tau$ | Equivalent time constant |
| $\gamma_i$ | Stability indices |
| $\gamma^*$ | Stability limits |
| $a_i$ | Coefficients of the characteristic polynomial |
| $K_{B0}$ | Zero-order term of $B_c(s)$ |
| $N_{pop}$ | Number of population streams created |
| $N_{Var}$ | Number of design variables |
| $N_{sr}$ | Number of rivers |
| $d$ | Distance between stream and river |
| $d_{max}$ | Small number to control search intensity |
| $UB$ | Upper bounds |
| $LB$ | Lower bounds |
| $J$ | Objective function |
| $K_{B0}$ | First term of $B(s)$ |
| $t_s$ | Settling time |
| $OS$ | Overshoot |
| $US$ | Undershoot |